\documentclass{moriond}

\def\be{\begin{equation}}
\def\ee{\end{equation}}
\def\bea{\begin{eqnarray}}
\def\eea{\end{eqnarray}}

\newcommand{\beq}{\begin{equation}}
\newcommand{\eeq}{\end{equation}}
\newcommand{\ta}{\mbox{{\boldmath
$\tau$}}}

\newcommand{\ro}{\mbox{{\boldmath
$\rho$}}}

\newcommand{\qb}{\mbox{{\bf
q}}}

\newcommand{\pt}{\mbox{{\bf
p}}_\perp}

\def\lsim{\mathrel{\rlap{\lower4pt\hbox{\hskip1pt$\sim$}}
    \raise1pt\hbox{$<$}}}         %less than or approx. symbol
\def\gsim{\mathrel{\rlap{\lower4pt\hbox{\hskip1pt$\sim$}}
    \raise1pt\hbox{$>$}}}         %greater than or approx. symbol

\def\lesssim{\mathrel{\rlap{\lower4pt\hbox{\hskip1pt$\sim$}}
    \raise1pt\hbox{$<$}}}         %less than or approx. symbol

\newcommand{\Photo}{}

\begin{document}
\vspace*{4cm}
\title{RADIATIVE QUARK $P_\perp$-BROADENING IN A QUARK-GLUON PLASMA AT RHIC
  AND LHC ENERGIES}

\author{ B.G. ZAKHAROV }

%\address{Department of Physics, Theoretical Physics, 1 Keble Road,\\
%Oxford OX1 3NP, England}
\address{L.D. Landau Institute for Theoretical Physics,
        GSP-1, 117940,\\ Kosygina Str. 2, 117334 Moscow, Russia}

\maketitle\abstracts{
We study the radiative correction to $p_\perp$-broadening of a fast quark 
in a quark-gluon plasma beyond the 
soft gluon approximation. 
We find that the radiative contribution to quark $\langle p_\perp^2\rangle$
for RHIC and LHC conditions is negative. 
}

\noindent {\bf 1}.
Parton transverse momentum broadening in a quark-gluon plasma (QGP)
is usually characterized by the transport coefficient $\hat{q}$ 
\cite{BDMPS1}: 
the mean squared transverse momentum of a fast parton passing through 
a uniform QGP of thickness $L$ is $\langle p_\perp^2\rangle=\hat{q}L$.
This is a leading order formula which includes only $p_\perp$-broadening
due to multiple scattering on the QGP constituents. 
The radiative processes can give an additional 
contribution to $p_\perp$-broadening. 
In the
soft gluon approximation
the radiative contribution
to $\langle p_\perp^2\rangle$ has been addressed 
in \cite{Wu_qhat-rad,Mueller_pt,Blaizot_pt}. In \cite{Mueller_pt}
it has been shown that radiative $p_\perp$-broadening 
is dominated by the double logarithmic contribution 
with 
$\langle p_{\perp}^2\rangle_{rad}\sim \frac{\alpha_sN_c\hat{q}L}{\pi}\ln^2(L/l_0)$
(where $l_0$ is about
the QGP Debye radius), and may be rather large. 

In this talk we consider
radiative $p_\perp$-broadening beyond the soft gluon and logarithmic 
approximations.
The analysis is based on the light-cone path integral (LCPI) 
\cite{LCPI1,LCPI_PT,Z_NP05} approach.
In the LCPI diagram technique of \cite{LCPI_PT}  
the spectrum of a $a\to bc$ process  in the Feynman variable $x$ 
and the transverse momentum  of particle $b$ is described by the
diagram Fig. 1a. For analysis of radiative $p_\perp$-broadening when $a=b$ 
one should also account for 
the virtual process $a\to bc\to a$ described by the diagram Fig.~1b.  
We perform calculations for $q\to qg$ process, i.e., for $a=b=q$ and $c=g$. 

\noindent {\bf 2}.
We consider a fast quark with energy
$E$ produced at $z=0$ (we choose the $z$-axis along
the initial quark momentum) traversing a uniform medium of thickness
$L$. 
We neglect collisional
energy loss (which is relatively small
\cite{Z_coll}), then the energy of the
final quark without gluon emission equals $E$,
for the two-parton final state the total energy also equals $E$. 
However, medium
modifies the relative fraction of the one-parton state and its transverse
momentum distribution, and for the two-parton channel medium modifies 
both the $x$ and the transverse momentum distributions.

In the approximation of single gluon emission
the radiative contribution to $\langle p_\perp^2\rangle$ reads
\beq
\langle p_\perp^2\rangle_{rad}=\int dx_q d\pt \pt^2
\left[\frac{dP}{dx_qd\pt}+\frac{d\tilde{P}}{dx_qd\pt}\right]\,,
\label{eq:10}
\eeq
where $\frac{dP}{dx_qd\pt}$ is the distribution for 
real splitting $q\to qg$ in the transverse momentum of the final quark
and its fractional longitudinal momentum $x_q$ (it corresponds to diagram 
Fig.~1a),
$\frac{d\tilde{P}}{dx_qd\pt}$ is the distribution
for the virtual process $q\to qg\to q$ (it corresponds to diagram Fig.~1b). 
In the latter case $x_q$ means 
the quark fractional momentum in the intermediate $qg$ system,
but $\pt$, as for the real process, corresponds to the final quark.  
The $x_q$-integration in (\ref{eq:10}) can equivalently be written
in terms of the gluon fractional momentum $x_g=1-x_q$.

\begin{figure}%[ht]
\begin{center}
\includegraphics[height=2.7cm]{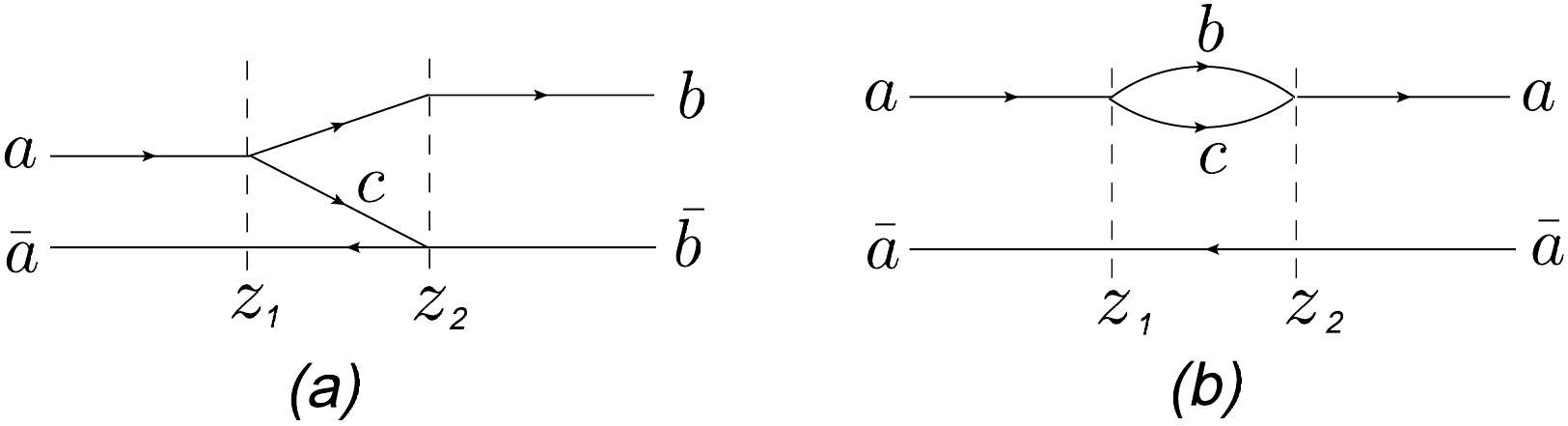}
\caption{\small Diagrammatic representation of ${dP}/{dx_bd\pt}$ ($a\to bc$
  process) (a) and of its virtual counterpart $d\tilde{P}/dx_bd\pt$
($a\to bc\to a$ process) (b). There are more two graphs
with interexchange of vertices between the upper and lower lines.
}
\end{center} 
\end{figure}

Let us consider the real splitting.
The distribution in the transverse momentum
and the longitudinal fractional momentum of the particle $b$
for $a\to bc$ transition 
corresponding to the graph of Fig.~1a has the form \cite{LCPI_PT}
\beq
\frac{dP}{dx_b d\pt}=\frac{1}{(2\pi)^{2}}
\int
d\ta_f\,\exp(-i\pt\ta_f)F(\ta_f)\,,
\label{eq:20}
\eeq
where
\bea
F(\ta_f)=
2\mbox{Re}
\int_{0}^{\infty} dz_{1} \int_{z_{1}}^{\infty} dz_{2}
\Phi_{f}(\ta_f,z_{2})
\hat{g}
K(\ro_{2},z_{2}|\ro_1,z_{1})
\Phi_{i}(\ta_i,z_{1})\Big|_{\ro_2=\ta_f, \ro_1=0}
\,,
\label{eq:30}
\eea
\beq
\Phi_{i}(\ta_i,z_{1})=
\exp\left[-\frac{\sigma_{a\bar{a}}(\ta_i)}{2}
\int_{0}^{z_{1}} dz\, n(z)\right],\,\,\,\,\,
\Phi_{f}(\ta_f,z_{2})=
\exp\left[-\frac{\sigma_{b\bar{b}}(\ta_f)}{2}
\int_{z_{2}}^{\infty} dz\, n(z)\right],
\label{eq:40}
\eeq
$\ta_i=x_b\ta_f$, 
$n(z)$ is the number density of the medium,
$\sigma_{a\bar{a}}$ and $\sigma_{b\bar{b}}$ are the dipole cross
sections for the $a\bar{a}$ and $b\bar{b}$ pairs,
$\hat{g}$ is the vertex operator,
$K$ is the Green function for the Hamiltonian
\beq
H=\frac{\qb^2+\epsilon^2}{2M}
-\frac{in(z)\sigma_{\bar{a}bc}(\ta_i,\ro)}{2}\,,
\label{eq:50}
\eeq
where $\qb=-i\partial/\partial \ro$, $M=E_{a}x_bx_c$, 
$\epsilon^2=m_{b}^{2}x_c+m_{c}^{2}x_b-m_{a}^{2}x_bx_c$
 with $x_c=1-x_b$,
and 
$\sigma_{\bar{a}bc}$ is the cross section for the 
three-body $\bar{a}bc$ system. The relative transverse parton positions 
for the $\bar{a}bc$ state read:
$\ro_{b\bar{a}}
=\ta_i+x_c\ro$,
$\ro_{c\bar{a}}
=\ta_i-x_b\ro$.
The vertex operator in (\ref{eq:30}) is
\beq
\hat{g}
=\frac{\alpha_s P^b_a(x_b)g(z_1)g(z_2)}{2 M^2}
\frac{\partial}{\partial\ro_1}\cdot
\frac{\partial}{\partial\ro_2}\,,
\label{eq:60}
\eeq
 where $P^b_a(x_b)$ is the $a\to b$  splitting function.
Because the $z$-integrations in (\ref{eq:30}) extend up to
infinity, and the adiabatically vanishing at $z\to \infty $ coupling 
$g(z)$ should be used. 
The three-body cross section 
can be written via 
the dipole cross section $\sigma_{q\bar{q}}$ for the $q\bar{q}$ system.
We will use the quadratic approximation 
$\sigma_{q\bar{q}}(\rho)=C\rho^2$ with $C=\hat{q}/2n$.
In this case the Hamiltonian (\ref{eq:50}) takes 
the oscillator form, and one can use analytical formula for the Green function.

To separate in (\ref{eq:30}) the contribution of the vacuum decay 
it is convenient to write the product 
$\Phi_{f}(\ta_f,z_{2})
\hat{g}
K(\ro_{2},z_{2}|\ro_1,z_{1})
\Phi_{i}(\ta_i,z_{1})
$
in the integrand on the right-hand side of (\ref{eq:30}) 
as (we denote $\hat{g} K$ as ${\cal K}$ and omit arguments for 
notational simplicity)
\beq
\Phi_f{\cal K}\Phi_i=\Phi_f({\cal{K}}-{\cal{K}}_0)\Phi_i+
(\Phi_f-1){\cal K}_0\Phi_i
+{\cal{K}}_0(\Phi_i-1)+{\cal K}_0\,,
\label{eq:70}
\eeq
where ${\cal K}_0=\hat{g}K_0$, and $K_0$ is the vacuum Green function.
The last term ${\cal K}_0$ in  (\ref{eq:70}) can be omitted because 
it does not contain medium effects.

The $\langle p_\perp^2\rangle_{rad}$ given by (\ref{eq:10}) 
may be written  via the Laplacian 
$\nabla^2$ at $\ta_f=0$  of the function $F$ and its counter part $\tilde{F}$ 
for the virtual diagram Fig.~1b. The result reads
%. The result can be written as
\beq
\langle p_\perp^2\rangle_{rad}=I_1+I_2+I_3\,,
\label{eq:80}
\eeq
\bea
I_1=2\mbox{Re}\int dx_q\int_{0}^L dz_1
\int_{0}^{\infty} d\Delta z
\nabla^2[{(\cal K}-{\cal K}_0)
-
(\tilde{{\cal K}}-\tilde{{\cal K}}_0)]\,,
\label{eq:90}
\eea
\bea
I_2=
2\mbox{Re}\int\!dx_q\int_{0}^L dz_1
\int_{0}^{\infty}\!d\Delta z
\left[
({\cal K}-{\cal K}_0)\nabla^2\Phi_i
-
(\tilde{{\cal K}}-\tilde{{\cal K}}_0)\nabla^2\tilde{\Phi}_i
\right]
\nonumber\\
=
-2\langle p_\perp^2\rangle_0
\mbox{Re}\int dx_q f(x_q)\int_{0}^L dz_1
\frac{z_1}{L}\int_{0}^{\infty}d\Delta z
({\cal K}-{\cal K}_0)\,,\,\,\,\,\,
\label{eq:100}
\eea
\bea
I_3=2\mbox{Re}\int dx_q\int_{0}^\infty dz_1\int_{0}^{\infty}d\Delta z
\left[{\cal K}_0\nabla^2\Phi_i-\tilde{{\cal K}}_0\nabla^2\tilde{\Phi}_i\right]
\nonumber\\
=-2\mbox{Re} \int dx_q f(x_q)
\int_{0}^\infty dz_1\int_{0}^{\infty} d\Delta z
{\cal K}_0\nabla^2\tilde{\Phi}_i\,\,\,\,\,\,\,
\label{eq:110}
\eea
with 
$
f(x_q)=1-x_q^2\,$,
and $\Delta z=z_2-z_1$.
In (\ref{eq:90})--(\ref{eq:110}) all functions in the integrands 
should be calculated at $\ta_f=0$ (as in (\ref{eq:70}), we omit arguments for simplicity). 
In (\ref{eq:100}), (\ref{eq:110}) we used 
that  at  $\ta_f=0$   
${\cal K}=\tilde{\cal K}$, ${\cal K}_0=\tilde{\cal K}_0$,
$\nabla^2\Phi_i=x_q^2\nabla^2\tilde{\Phi}_i$,
and  $\nabla^2\tilde{\Phi}_i$ equals $-\langle p_\perp^2\rangle_0z_1/L$,
where 
$\langle p_\perp^2\rangle_0$ 
corresponds to nonradiative $p_\perp$-broadening.
The integration over $z_{1}$ in (\ref{eq:110})
is unconstrained,
and should be performed for an adiabatically vanishing 
coupling $g(z)$ in (\ref{eq:60}). We use $g(z)\propto \exp(-\delta z)$.
Taking the limit $\delta\to 0$ after calculations for a finite $\delta$ 
 we obtain for $I_3$
\beq
I_3=-\langle p_\perp^2\rangle_0 \int dx_q
f(x_q)\frac{dP_0}{dx_q}\,,
\label{eq:120}
\eeq
where 
\beq
\frac{dP_0}{dx_q}=\int d\pt
\frac{dP_0}{dx_q d\pt}
\label{eq:130}
\eeq
is the $\pt$-integrated
vacuum spectrum.
The $\pt$-integral 
in (\ref{eq:130}) is logarithmically divergent.
This occurs because we work
in the small angle approximation \cite{LCPI_PT}, which 
ignores the kinematic limits.  
We regulate (\ref{eq:130}) by restricting the integration region to 
$p_\perp\!<p_\perp^{max}=E\mbox{min}(x_q,(1-x_q))$.

The $\Delta z$-integral in (\ref{eq:90}) is also logarithmically 
divergent, because
the integrand is $\propto 1/\Delta z$ as $\Delta z\to 0$. 
It is reasonable to regulate the $\Delta z$-integral
in (\ref{eq:90}) by using the lower limit $\Delta z\sim 1/m_D$ 
This prescription has been used in \cite{Mueller_pt} for calculation
in the logarithmic approximation
of the contribution corresponding to our $I_1$  (\ref{eq:90}). 
The contributions from 
$I_2$ and $I_3$ terms
have not been included in \cite{Mueller_pt}. 

\noindent {\bf 3}.
In numerical calculations we use the quasiparticle masses $m_{q}=300$ and $m_{g}=400$ MeV  
\cite{LH}, that have been used in our previous 
analyses \cite{RAA13,RPP14}
of the RHIC and LHC data on the nuclear modification factor 
$R_{AA}$. 
The calculations of \cite{RAA13,RPP14} have been performed
for a more sophisticated model with running $\alpha_s$ 
for the QGP with Bjorken's longitudinal expansion,
which corresponds to $\hat{q}\propto 1/\tau$. 
In the present work, as in \cite{Mueller_pt}, we use constant 
$\hat{q}$ and $\alpha_s$.
To make our estimates more realistic 
we adjusted $\hat{q}$ to reproduce the quark
energy loss $\Delta E$ for running $\alpha_s$
in the model of \cite{RPP14} with the Debye mass from the lattice
calculations \cite{Bielefeld_Md}. 
We obtained $\hat{q}\approx 0.12$ GeV$^3$ at $E=30$ GeV 
for Au+Au collisions
at $\sqrt{s}=0.2$ TeV and 
$\hat{q}\approx 0.14$ GeV$^3$ at $E=100$ GeV for Pb+Pb collisions
at $\sqrt{s}=2.76$ TeV.
As in \cite{Mueller_pt}, we take $\alpha_s=1/3$ and $L=5$ fm.

We have taken into account that the transport coefficient that describes
the Glauber factors $\Phi_i$ and $\tilde{\Phi}_i$ in the formulas
for $I_{2,3}$ may differ from $\hat{q}$ that controls the Green functions
in $I_{1,2}$. For the Glauber factors $\hat{q}$ should be calculated at 
the energy of the initial quark $E$, 
but for the Green functions it is reasonable
to use the transport coefficient at the typical energy of the radiated gluon
$\bar{\omega}$.
The above adjusted values of $\hat{q}$
correspond just to the transport 
coefficients for gluons with energy $\sim\bar{\omega}$.
We denote the transport coefficient for the Glauber factors $\hat{q}'$.
Since $E\gg \bar{\omega}$, the ratio $r=\hat{q}'/\hat{q}$ 
may differ significantly from unity.
Using the Debye mass from \cite{Bielefeld_Md} and
running $\alpha_s$ parametrized as in our previous 
jet quenching analysis 
\cite{RPP14} we obtained 
$
r\approx 1.94(2.13)
$
at $E=30(100)$ GeV for quark jets for RHIC(LHC) conditions.

In numerical calculations in (\ref{eq:90})--(\ref{eq:110}) 
we integrate over $x_q$ from $x^{min}_q=m_q/E$
to $x^{max}_q=1-m_g/E$.
As in \cite{Mueller_pt}, for the cutoff in the $\Delta z$-integration 
we use $\Delta z_{min}=1/m$ with $m=300$ MeV. 
With these parameters 
%for 
%the three terms in (\ref{eq:80})
we obtained at $E=30$ GeV for the RHIC conditions
\beq
[I_1,I_2,I_3]/\langle p_\perp^2\rangle_{0}\approx
[0.417/r,-0.213,-0.601]\,,
\label{eq:140}
\eeq
and at $E=100$ GeV for the LHC conditions
\beq
[I_1,I_2,I_3]/\langle p_\perp^2\rangle_{0}\approx
[0.823/r,-0.107,-0.908]\,.
\label{eq:150}
\eeq
From (\ref{eq:80}), (\ref{eq:140}) and (\ref{eq:150})
for our RHIC(LHC) versions we obtain 
\beq
\langle p_\perp^2\rangle_{rad}/
\langle p_\perp^2\rangle_{0}
\approx -0.598 (-0.629)\,,\,\,
r=  1.94(2.13)\,.
\label{eq:160}
\eeq
And if we take $r=1$ 
\beq
\langle p_\perp^2\rangle_{rad}/
\langle p_\perp^2\rangle_{0}
\approx -0.397(-0.192)\,,\,\,
r=1(1)\,.
\label{eq:170}
\eeq
Thus, in all the cases
the radiative contribution is negative. We have checked that
under  variations
of parton masses by a factor of $\sim 2$ 
the value of $\langle p_\perp^2\rangle_{rad}$ remains negative.   
Our predictions  differ drastically from 
$\langle p_\perp^2\rangle_{rad}\approx 0.75\hat{q}L$
obtained in \cite{Mueller_pt}.
The negative values of 
$\langle p_\perp^2\rangle_{rad}$ in our calculations are due to 
large negative values of $I_{2,3}$.
Since these terms 
have not been accounted for in \cite{Mueller_pt},  
it is interesting to compare prediction of \cite{Mueller_pt} 
with our results for $I_1$ term alone.
From (\ref{eq:140}) and (\ref{eq:150}) one can see that 
our $\left.\langle p_\perp^2\rangle_{rad}\right|_{I_1}$
agrees qualitatively with $\langle p_\perp^2\rangle_{rad}$
from \cite{Mueller_pt}.

\noindent {\bf 4}.
In summary, we have studied within the LCPI \cite{LCPI1,LCPI_PT,Z_NP05} 
approach the radiative contribution to $p_\perp$-broadening 
of fast quarks in the QGP. The analyses is performed beyond the 
soft gluon approximation. We have found that 
for  RHIC and LHC  conditions
the radiative contribution to quark $\langle p_\perp^2\rangle$
may be negative. 
This seems to be supported by
the recent STAR measurement of the hadron-jet correlations \cite{STAR1},
in which no  evidence  for large-angle jet scattering in
the medium has been found.\\  

This work was partly supported by the RFBR grant 
18-02-40069mega.

%\subsection{Producing the Hard Copy}\label{subsec:prod}

%\section*{Acknowledgments}

\end{document}